\documentclass{elsart}
\usepackage{amssymb}   
\usepackage{graphicx}
\begin{document}

\setlength\paperwidth{8.5in}
\setlength\paperheight{11in}
\setlength{\pdfpagewidth}{\paperwidth}
\setlength{\pdfpageheight}{\paperheight}
\interfootnotelinepenalty=100000

\begin{frontmatter}

\title{Methods for point source analysis in high energy neutrino telescopes}

\author[label1]{Jim Braun\corauthref{cor}}, 
\ead{jim.braun@icecube.wisc.edu}
\author[label1]{Jon Dumm}, 
\author[label1]{Francesco De Palma}, 
\author[label1]{Chad Finley}, 
\author[label1]{Albrecht Karle}, 
\author[label1,label4]{Teresa Montaruli}
\address[label1]{University of Wisconsin, Chamberlin Hall, Madison, Wisconsin 53706, USA}
\address[label4]{On leave of absence from 
Universit\`a di Bari and INFN, 70126 Bari, Italy}
\corauth[cor]{Corresponding author.}

\begin{abstract}
Neutrino telescopes are moving steadily toward the goal of detecting astrophysical
neutrinos from the most powerful galactic and extragalactic sources.
Here we describe analysis methods to search for high energy point-like neutrino sources using detectors
deep in the ice or sea. We simulate an ideal cubic kilometer detector based on real world performance
of existing detectors such as AMANDA, IceCube, and ANTARES.
An unbinned likelihood ratio method is applied, making use of the point spread function and energy distribution of simulated neutrino
signal events to separate them from the background of atmospheric neutrinos produced by cosmic ray showers.  The unbinned point source analyses are
shown to perform better than binned searches and, depending on the source spectral index,
the use of energy information is shown to improve discovery potential by almost a factor of two.

\end{abstract}
\end{frontmatter}

\section{Introduction}
\label{Sec1}

With the construction of IceCube at the South Pole \cite{IceCube} and of ANTARES in the Mediterranean Sea \cite{ANTARES},
together with existing R\&D programs for a cubic kilometer array at these latitudes \cite{km3net,NEMO}, neutrino astronomy
is entering a very promising era. IceCube, when complete in 2011, will consist of up to 80 strings on a hexagonal grid with 124 m spacing,
with each string holding 60 optical modules vertically spaced by 17 m. The instrumented part of the strings is
deployed between 1450 and 2450 m deep in the ice. The experiment profits from experience acquired with the AMANDA detector, 
taking data since 1996 and completed in 2000 with 19 strings containing 677 total optical modules between
1500 and 2000 m below the ice surface.
Studies on the optimal configuration for a cubic kilometer array in the Mediterranean are ongoing, and the amount of
photomultipliers (PMTs) generally considered is somewhat larger than for IceCube, around 6000 \cite{distefano}
and up to about 9000 \cite{Carr}.

The community is refining methods to detect low statistics signals amongst large backgrounds. The
expected background from atmospheric neutrinos in a cubic kilometer detector is of the order of 50 000 upgoing events
per year after selection criteria guaranteeing good angular resolution and rejection of misreconstructed
cosmic ray muons.  Point-like signals of
few events need to be singled out among this large number of background events. Two features distinguish
signal from the background:
\begin{itemize}
\item The angular distribution. The signal would cluster around the direction of the neutrino source (assumed here to be
point-like) with a spread depending on the detector angular resolution.
Angular resolution is limited by detector geometry and by the propagation characteristics
of light in the medium, specifically by photon scattering and absorption. The pointing accuracy for astrophysical sources
is also limited by the kinematic angle between the parent neutrino and the muon.
\smallskip
\item The energy distribution. The differential energy spectrum of the signal expected from Fermi acceleration
mechanisms is close to E$^{-2}$, harder than that of atmospheric neutrinos due to the showering process in the atmosphere.
The differential spectrum of atmospheric neutrinos approximately follows a power law of E$^{-3.7}$ above 100 GeV.
\end{itemize}
Other signatures may be used, including time dependencies of emissions measured in other detectors such as correlations
with gamma ray bursts or TeV gamma ray flares.  However, we focus on steady emissions of neutrinos with time.
The methods we have implemented exploit the two features listed above. We show that unbinned methods based on
the likelihood ratio hypothesis test perform better than methods based on angular bins, 
and we show that the introduction of energy dependent information, e.g. the number of hit
PMTs, helps discriminate signal and allows an energy spectrum reconstruction even when few events are detected on top of the
background. Other unbinned methods have been studied and developed by other authors \cite{till2,aart,finley,aguilar}.

Sec.~\ref{Sec2} describes the simulation we use to generate realistic samples of signal and background events
in a cubic kilometer detector.
Sec.~\ref{Sec3} describes the unbinned method based on a likelihood ratio analysis, comparing a signal plus
background hypothesis to a background only hypothesis. This method has been applied to IceCube data for 9 strings and
to 2005-6 data of the AMANDA-II detector ~\cite{chadicrc07,jimicrc07}. In Sec.~\ref{Sec4} we 
describe the performance of the methods and show results in terms of discovery potential. We also emphasize the
importance of using energy related information to increase the discovery potential and show the ability to determine
the spectral index of the neutrino source.
The method is then compared to a search using angular bins, more traditionally applied in neutrino astronomy
\cite{macro,amanda5,SuperKamiokande}.

\section{Simulation of a data sample of atmospheric neutrino background and point source signal}
\label{Sec2}

We wish to compare several neutrino point source search methods and draw general conclusions on the discovery potential
for cubic kilometer scale neutrino telescopes under construction in the ice and under study in the sea water.
We have performed a realistic simulation of both atmospheric neutrino events and signal events from an astrophysical
neutrino point-like source.  This is accomplished by a detailed detector volume
simulation with simplifications expected to have negligible impact on the comparison.

\begin{figure}\begin{center}
\mbox{\includegraphics[width=4in]{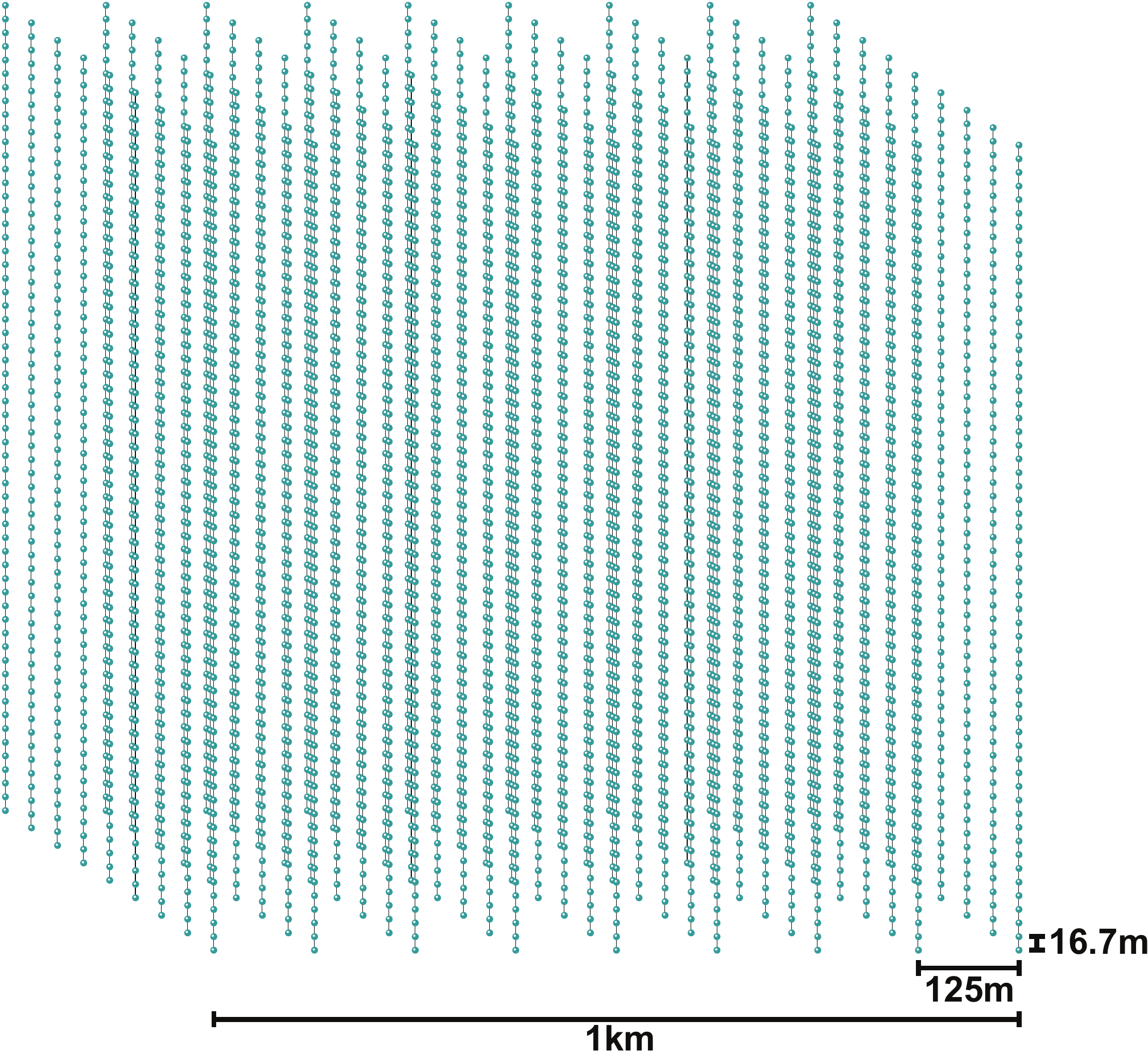}}
\caption{\label{Fig:Array} Simulated detector consisting of 81 strings, each containing 60 optical modules.}
\end{center}\end{figure}

The simulated detector 
consists of 4860 optical modules arranged in 81 strings.  The strings are
evenly distributed on a square $9 \times 9$ grid with 125 meters separating nearest neighbors,
and the modules on each string are vertically separated by 16.67 m, shown in Fig.~\ref{Fig:Array}.
The modules are simulated as containing a downward looking 10 inch photomultiplier with 20\% quantum efficiency.
A multiplicity trigger requires that at minimum 14 modules register a photon from an event.  We neglect
trigger time windows and photon hits from photomultiplier dark noise, since hits unrelated to
the track can presumably be removed with coincidence requirements, filtering strategies, and topological cuts.
We assume the detector is located at the South Pole. A different location would imply a different visibility
of sources in the sky, since at the South Pole half of the sky is always visible while the other half is inaccessible.
At other latitudes the detector is always blind to less than one half of the sky, always sensitive to a region of
similar size in the opposite hemisphere, and sensitive to the remainder of the sky for a fraction of the day.

We have prepared an algorithm to simulate the muon neutrino interactions in a volume and 
propagate the secondary muon in the ice, producing the light detected by the PMTs.
For neutrino generation, we use an updated version of the simulation described in \cite{teresa}.
We use the more recent  CTEQ6 \cite{CTEQ6} structure functions
for the deep inelastic cross section of muon neutrinos and simulate the Earth density profile in \cite{gandhi}
to account for the absorption of high  energy neutrinos. The muon is propagated to the instrumented region using the MUM
propagation code \cite{igor}. We model the muon energy loss within the detector as dE/dx = a + bE, where
a = 0.268 GeV/m accounts for ionization energy losses, and b = 0.00047~m$^{-1}$
for bremsstrahlung, pair production, and 
photonuclear interactions \cite{MMC}. The number of
PMTs recording light, or `hit', increases with the energy loss rate, and thus the energy, of the muon.
Since the energy dependent processes are stochastic, treating them as continuous
may overestimate the number of modules hit by a small amount at PeV energies.
However, high energy muons where this effect is significant are very bright and tend to produce
photon hits in many more modules than the trigger threshold of 14, so we conclude that the impact on event triggering is negligible.

\begin{figure}\begin{center}
\mbox{\includegraphics[width=4in]{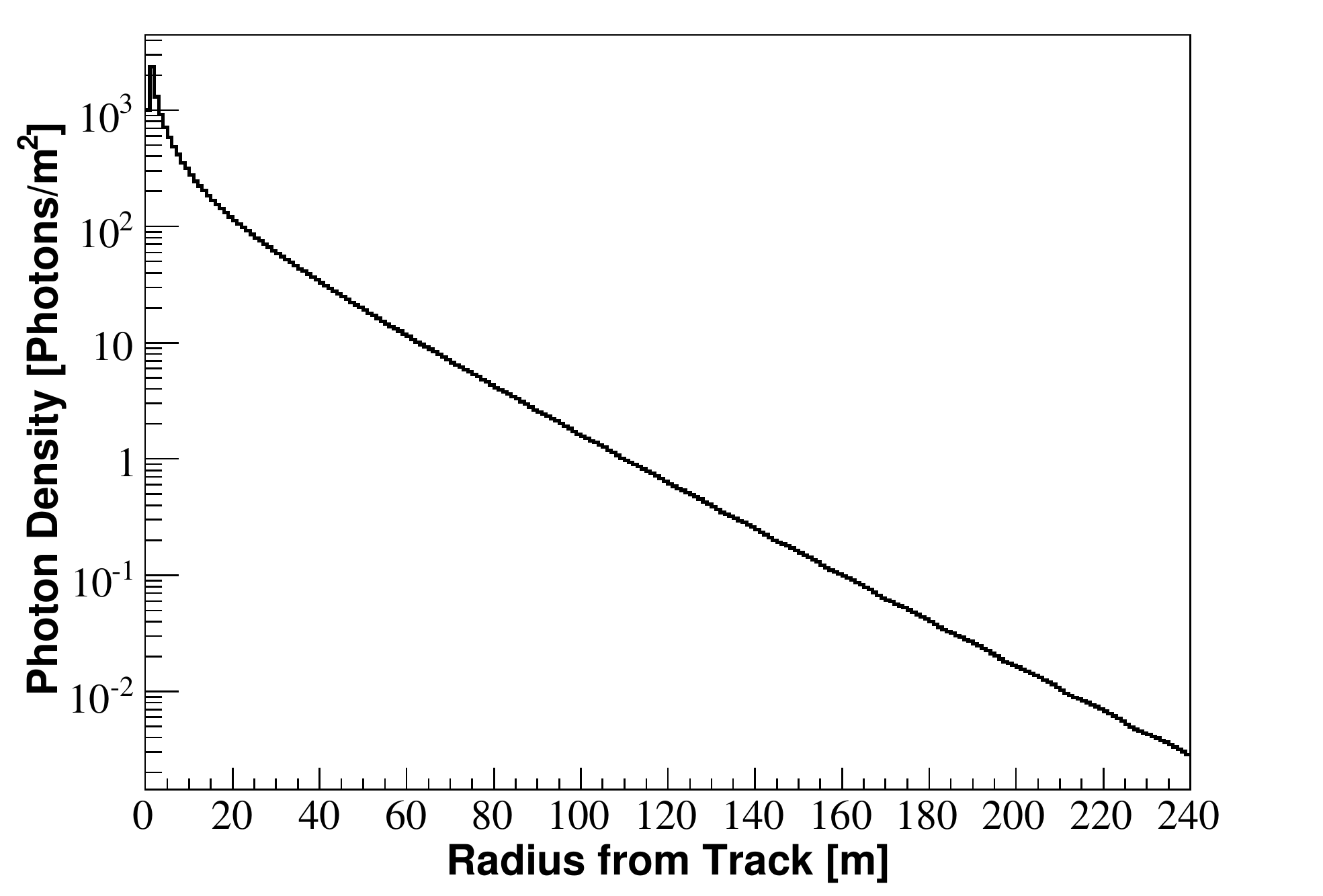}}
\caption{\label{Fig:photondensity} Photon density as a function of the radius from a minimum ionizing muon track in ice with an
effective scattering length of 21 meters and absorption length of 120 meters.}
\end{center}\end{figure}

Photon propagation in the detector medium has a significant impact on the detector response.
We assume a homogeneous detector medium in which the photon density at the sensor
can be described as a simple function with respect to the distance of the muon track.
Photons are propagated with an effective scattering length of 21 m and an absorption length of 120 m,
which are typical values for South Pole ice \cite{icepaper}.
Fig. \ref{Fig:photondensity} shows the assumed photon density for a minimum ionizing muon as a function of distance
for the case of ice.  For a muon with arbitrary energy loss dE/dx, the photon density is scaled by the equivalent energy
loss of a minimum ionizing muon, i.e. scaled by dE/dx / (0.268 GeV/m).  The number of photoelectrons recorded
from each PMT is a Poisson random variable with mean equal to the product of this photon density, PMT photocathode area,
and PMT quantum efficiency.  For each PMT, a hit is determined by randomly sampling this Poisson probability
of observing at least one photoelectron.

\begin{figure}\begin{center}
\mbox{\includegraphics[width=4in]{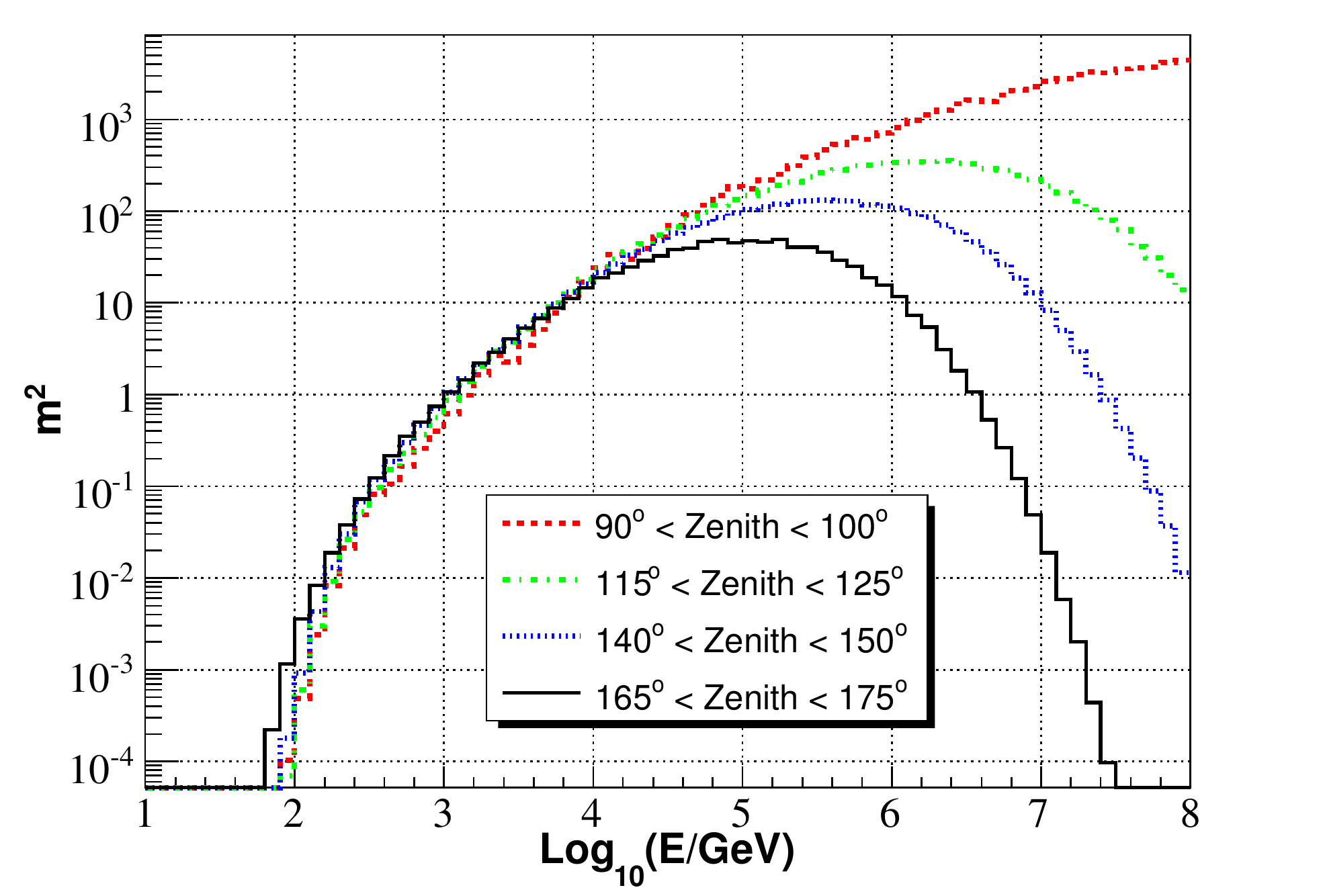}} 
\caption{\label{Fig:Aeff} Neutrino effective area at trigger level for several zenith bands.}
\end{center}\end{figure}

We have generated $10^{11}$ upgoing neutrino events isotropically entering the Earth.
The events are drawn from an E$^{-1.4}$ energy spectrum between 10 and
$10^{9}$ GeV. This hard spectrum allows efficient generation of high energy events.
Events passing the trigger threshold of $\ge 14$ hit PMTs are kept,
resulting in the neutrino effective area shown in Fig.~\ref{Fig:Aeff}.  
The neutrino effective area represents the equivalent detector area for a hypothetical instrument with
100\% efficiency for detecting the passage of neutrinos.  The effective area is much smaller than the dimensions of the detector
due to the small cross section of neutrino interaction and, for larger zenith angles and high energies, neutrino absorption on transit through the Earth.
It is a useful parameter for determining event rates and making comparisons between experiments.
The event rate for a neutrino model predicting
a flux $\frac{d\Phi}{dE_{\nu}d\Omega_{\nu}}$ is given by
\begin{equation}\frac{dN_{\mu}}{dt} = \int\int dE_{\nu}d\Omega_{\nu} A^{eff}_{\nu}(E_{\nu}, \Omega_{\nu})\frac{d\Phi}{dE_{\nu}d\Omega_{\nu}}.
\end{equation}

\begin{figure}\begin{center}
\begin{tabular}{cc}
\mbox{\includegraphics[width=2.57in]{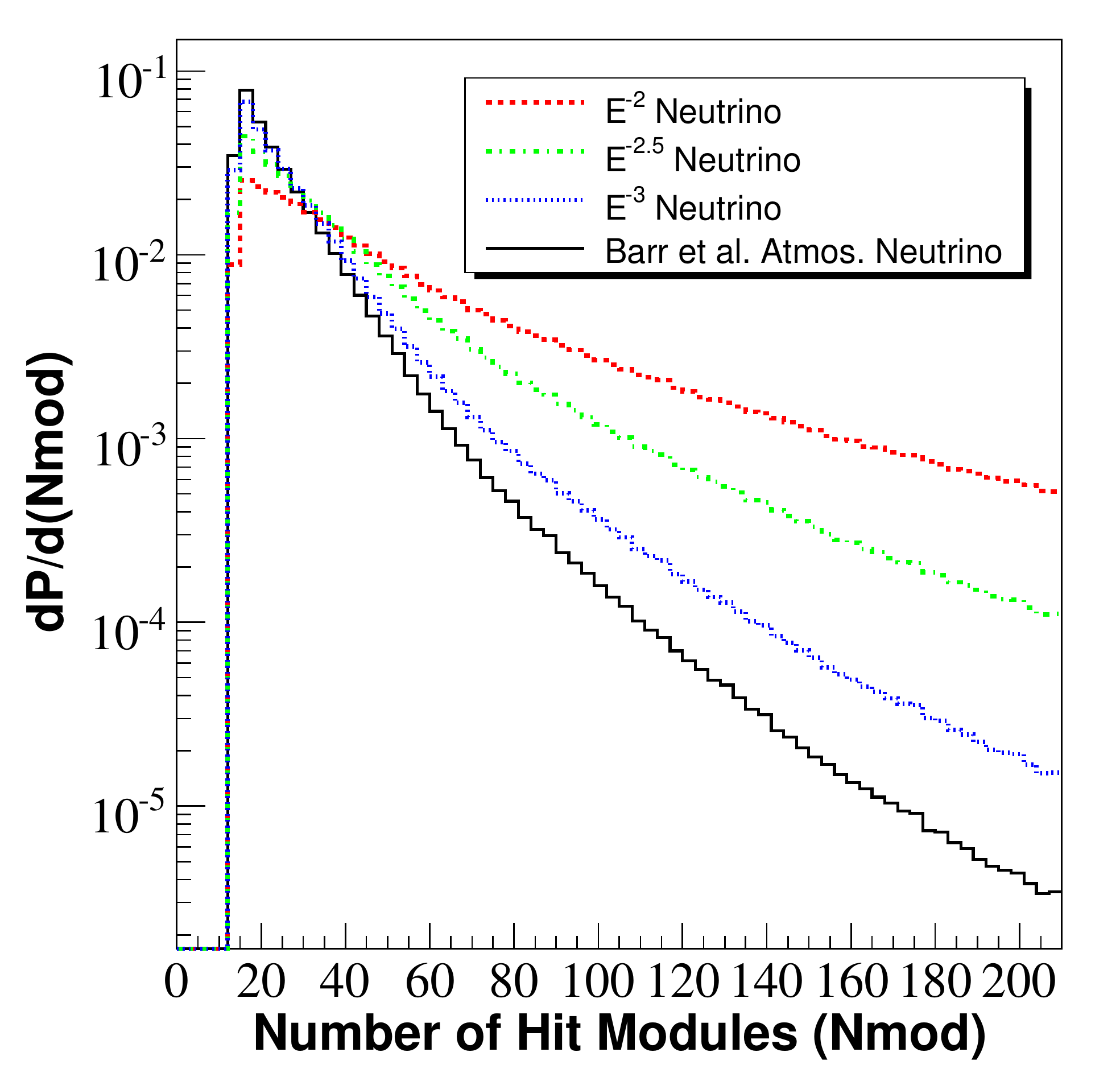}}
\mbox{\includegraphics[width=2.57in]{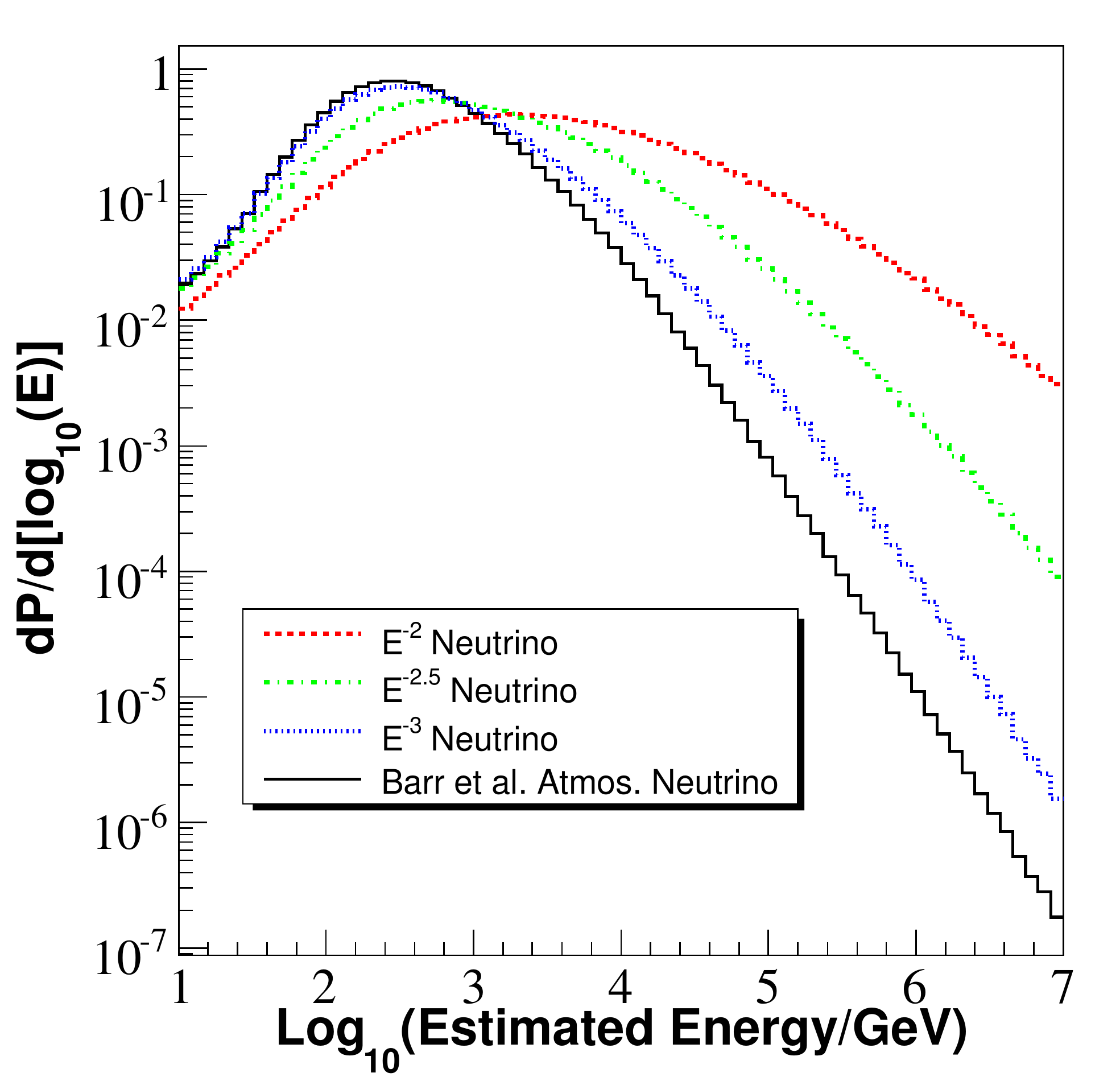}}
\end{tabular}
\caption{\label{Fig:E_est} Number of hit modules distribution (left) and reconstructed muon energy distribution (right)
for several neutrino spectra.}
\end{center}\end{figure}

Using the atmospheric neutrino flux of Barr et al.
\cite{bartol}, we find approximately 134 000 atmospheric neutrino events per year at trigger level with a median energy of 670 GeV
and a slight density dependence on zenith angle, and thus declination, of $\pm$15\%.
This number of events will be reduced however by topological and reconstruction cuts necessary to eliminate the large
background of misreconstructed downgoing
muons from cosmic ray air showers. We estimate 50\% of atmospheric events will be lost to reject this background;
thus we choose 67 000 events from the atmospheric neutrino sample for our data set, roughly representing a year of data.
Cuts applied to a real data sample may have some small energy dependence, i.e. efficiency may drop for low energy events,
but we do not consider this effect.
We generate astrophysical signal neutrinos similarly; however, they are simulated at a fixed declination and weighted differently than
atmospheric neutrinos. We consider for the signal a power law neutrino spectrum with spectral indices ranging from
1.5 to 4.0. The number of hit modules distribution for atmospheric neutrinos and several
signal spectral indices is shown in Fig.~\ref{Fig:E_est}.

Angular reconstruction errors for a cubic kilometer detector in ice are shown to be $\sim$0.7$^{\circ}$ for neutrinos of energies
$\gtrsim 1$ TeV \cite{IceCube}.
We simulate this reconstruction error by adding to the muon direction a space angle error randomly sampled from a polar
Gaussian distribution with standard deviation 0.7$^{\circ}$.  Considering both this reconstruction error and the neutrino-muon vertex angle,
the resulting median angular resolution for an E$^{-2}$ signal is 0.86$^{\circ}$ and is larger for softer spectra.
A comparison to a detector with angular resolution 0.2$^{\circ}$, which may be achieved with a detector in sea water,
is shown in Sec.~\ref{Sec4}.

Muon energy reconstruction errors in neutrino telescopes are typically of the order of 0.3 in $\log_{10}$E$_{\nu}$ above a few TeV 
\cite{amanda_reco,dimajuande,romeyer},
limited by the stochastic nature of muon energy losses. To simulate this effect, we assign to each event a
reconstructed error sampled from a Gaussian distribution of width 0.3 in the logarithm of energy.  
The reconstructed energy for atmospheric neutrino events and several power law spectra is shown in Fig.~\ref{Fig:E_est}.  
Energy resolution degrades below a few TeV; thus the energy resolution we assign to such events is unrealistically accurate. However, the
power to detect astrophysical sources resides in the ability to discriminate high energy neutrinos from lower energy atmospheric
neutrino background, so we conclude overestimating the energy resolution of such lower energy events does not affect the result significantly.

\section{The unbinned likelihood ratio method}
\label{Sec3}

In the context of a muon neutrino point source search, the data from a neutrino telescope consists
of a set of muon events spread over the sky, each with reconstructed direction (declination and right ascension),
energy, and time.  While event time is useful in searches for short neutrino bursts or periodic neutrino
emission, we focus on searches for continuous neutrino emission and disregard event time.
The vast majority of events are muons produced by atmospheric neutrinos.  At any
celestial direction, the data can be modeled by two hypotheses:
\begin{itemize}
\item $H_0$:  The data consists solely of background atmospheric neutrino events.
\item $H_S$:  The data consists of atmospheric neutrino events as well as astrophysical neutrino
events produced by a source with some strength and energy spectrum.
\end{itemize}
The likelihood of obtaining the data given each hypothesis is calculable,
and the ratio of likelihoods, or equivalently the log of the likelihood ratio, serves as a powerful
test.  We define our test statistic

\begin{equation}
\lambda = -2\cdot log\Bigg[\frac{P(Data|H_0)}{P(Data|H_S)}\Bigg].
\end{equation}

Larger values of $\lambda$ indicate the data is less compatible with the background hypothesis $H_0$.
The probability density functions $P(Data|H_0)$ and $P(Data|H_S)$ are calculated using knowledge of the spatial and
energy distribution of background and astrophysical neutrino events.

Suppose we wish to test the existence of a source at a known direction $\vec x_s$ using a set of
data events, each with reconstructed direction $\vec x_i$ and reconstructed energy $E_i$.  Events reconstructed
outside a declination band centered at $\vec x_s$ with a width several times the detector resolution are unlikely
to have been produced by a source at $\vec x_s$ and can be disregarded, leaving $N$ events in the band.  Each event
inside the band is assigned a source probability density corresponding to the probability of the event belonging
to a source at $\vec x_s$.  The source is assumed to emit neutrinos according to an E$^{-\gamma}$ power law
energy spectrum.  The source probability density is the product of a spatial density function describing the potential
of an event reconstructed with direction $\vec x_i$ to have true direction $\vec x_s$ and the probability of observing reconstructed
muon energy $E_i$ given source spectral index $\gamma$:

\begin{equation}
\mathcal{S}_{i}(\vec x_i, \vec x_s, E_i, \gamma) = \mathcal{N}(\vec x_i|\vec x_s)\cdot \int_{E_{\nu}} P(E_i|E_{\nu})P(E_{\nu}|\gamma)dE_{\nu} \, .
\end{equation}

The spatial probability density component $\mathcal{N}(\vec x_i|\vec x_s)$ can be obtained directly from event reconstruction
when maximum likelihood reconstruction techniques are used \cite{till}.  The structure of the reconstruction likelihood
near the most likely direction $\vec x_i$ provides an event by event estimate of reconstruction uncertainty and yields a Gaussian
spatial probability density profile:

\begin{equation}
\mathcal{N}_i(\vec x_i|\vec x_s) = \frac{1}{2\pi \sigma^2}e^{-\frac{|\vec x_i - \vec x_s|^2}{2\sigma^2}},
\end{equation}

where $|\vec x_i - \vec x_s|$ is the space angle difference between source and reconstructed event directions, and $\sigma$ is
the reconstruction error estimate.  The vertex angle between the neutrino and muon is neglected in $\mathcal{N}_i(\vec x_i|\vec x_s)$.
This angle is negligible compared to reconstruction error for cubic kilometer neutrino telescopes in ice with resolution
$\sim$0.7$^{\circ}$ and high energy threshold of $\sim$100 GeV.  Addition of the vertex angle error is discussed in section
\ref{Sec4} for cubic kilometer telescopes in sea water with resolution $\sim$0.2$^{\circ}$.  The integral over
$E_{\nu}$ is precomputed from detector Monte Carlo.  The resulting tables of $P(E_i|\gamma)$ for 1.0 $< \gamma <$ 4.0, shown
for several indices in Fig.~\ref{Fig:E_est}, are computed in steps of 0.01 in $\gamma$, interpolated linearly, and stored
for reference.\footnote{For maximum accuracy, one may wish to tabulate $P(E_i|\gamma)$ with respect to zenith angle as well; however, we omit
this step and achieve good results.}
The resulting source probability density is

\begin{equation}
\mathcal{S}_{i}(\vec x_i, \vec x_s, E_i, \gamma) = \frac{1}{2\pi \sigma^2}e^{-\frac{|\vec x_i - \vec x_s|^2}{2\sigma^2}}P(E_i|\gamma)
\end{equation}

and has value unity when integrated over solid angle and $E_i$.  The slight declination dependence of the atmospheric neutrino background
can be neglected over the width of the band.  The background probability density again depends on event energy and is then

\begin{equation}
\mathcal{B}_{i} = \frac{P(E_i|\phi_{atm})}{\Omega_{band}}.
\end{equation}

The probability density $P(E_i|\phi_{atm})$ is precalculated as above assuming the atmospheric neutrino flux of Barr et al.
\cite{bartol}.  We assume the background is purely atmospheric neutrinos.  If the background contains another component, for example
high energy muons from cosmic ray air showers, the energy component of the background density should be modified accordingly: 
$P(E_i|\phi_{atm} + \phi_{Bkgd})$.  The source and background densities are combined, and the likelihood is evaluated over all
events in the band:

\begin{equation}
\mathcal{L}(\vec x_s, n_s, \gamma) = \prod_{N} \Bigg(\frac{n_s}{N}\mathcal{S}_{i} + (1 - \frac{n_s}{N})\mathcal{B}_i \Bigg)
\end{equation}

where $n_s$ describes the number of signal events present in the band.%CF%

The fraction of signal events $n_s$ as well
as the source spectral index $\gamma$ are not known and must be determined by maximizing the likelihood $\mathcal{L}$.
This is done by minimizing the quantity $-log(\mathcal{L})$ using the MIGRAD minimizer available in MINUIT \cite{minuit}
with respect to the unknown quantities $n_s$ and $\gamma$ and obtaining the best value of each parameter, $\hat{n}_s$ and
$\hat{\gamma}$.  The minimization procedure also incorporates a penalty factor\footnote{The extension of $\gamma$ is limited above 2.7 during
the minimization by a Gaussian likelihood penalty with $\sigma$=0.2 in spectral index.  This improves the discovery potential for hard astrophysical
source spectra while remaining comparable to an unbinned method without energy for source spectra as soft as $\sim$3.2 (see Fig. \ref{Fig:Probvspec}).}
for $\gamma >$ 2.7 to better discriminate astrophysical sources with hard spectral indices from the atmospheric background with $\gamma \sim$3.7.
The original hypotheses can be written in terms of $\mathcal{L}$: $P(Data|H_0) = \mathcal{L}(\vec x_s, 0)$
and $P(Data|H_S) = \mathcal{L}(\vec x_s, \hat{n}_s, \hat{\gamma})$.  The test statistic is 

\begin{equation}
\lambda = -2 \cdot sign(\hat{n}_s) \cdot log\Bigg[\frac{\mathcal{L}(\vec x_s, 0)}{\mathcal{L}(\vec x_s, \hat{n}_s, \hat{\gamma})}\Bigg].
\end{equation}

Ignoring the factor $sign(\hat{n}_s)$, the test statistic $\lambda$ is never negative since $\mathcal{L}(\vec x_s, 0)$ is contained in the range of
$\mathcal{L}(\vec x_s, n_s, \gamma)$, of which $\mathcal{L}(\vec x_s, \hat{n}_s, \hat{\gamma})$ is the maximum.
A downward fluctuation of the background may occur at $\vec x_s$ which would be fit best as a source with
negative number of events and a negative value of $\hat{n}_s$.  Such a downward fluctuation also would have a large
value of $\lambda$, so $sign(\hat{n}_s)$ is used to separate negative and positive excesses.  A full sky
search is a simple extension of this single point search method and can be accomplished by performing the
search on a grid of locations covering the sky.  Finally, while it is preferable to use an energy estimation to maximize the power
to discriminate astrophysical neutrinos from the background, it is possible to do the search without this information.
The energy dependent terms are removed (i.e., set to one) from the signal and background probability densities,

\begin{equation}
\mathcal{S}_{i}(\vec x_i, \vec x_s) = \frac{1}{2\pi \sigma^2}e^{-\frac{|\vec x_i - \vec x_s|^2}{2\sigma^2}}
\end{equation}
\begin{equation}
\mathcal{B}_{i} = \frac{1}{\Omega_{band}}
\end{equation}

resulting in simpler expressions for $\mathcal{L}$.  The quantity $-log(\mathcal{L})$ is minimized with respect to $n_s$, and
$P(Data|H_0) = \mathcal{L}(\vec x_s, 0)$ and $P(Data|H_S) = \mathcal{L}(\vec x_s, \hat{n}_s)$.

\section{Results}
\label{Sec4}

We apply the likelihood ratio method to data consisting of the 67 000 background events described
in section \ref{Sec2} and an added source at declination 48$^{\circ}$.  A grid of simulated source strengths
and spectral indices is used, with source strength spanning 0 - 100 signal events added to the sample
and spectral index $\gamma$ spanning 1.0 - 3.9 in increments of 0.1, resulting in a total of 3000 simulated combinations
of source strength and spectral index.  
For each combination, 10 000 experiments are done and
the value of $\lambda$ is recorded for each.  10$^7$ trials are performed with background alone to evaluate
the significance of observed values of $\lambda$.  The method is compared against a binned search with
a circular bin centered at the source location.  The bin radius is optimized by minimizing the number of E$^{-2}$ signal
events necessary to achieve 5$\sigma$ significance in 90\% of experiments.  The optimal radius is found to be
1.35$^{\circ}$, with a signal efficiency of 80\% and background expectation of $\sim$16 events/bin.  Also, the
method is compared against itself using only spatial information and neglecting event energy.  An identical
grid of simulated source strengths and spectral indices is used for both the binned method and likelihood
method without energy.

\begin{figure}\begin{center}
\mbox{\includegraphics[width=5.4in]{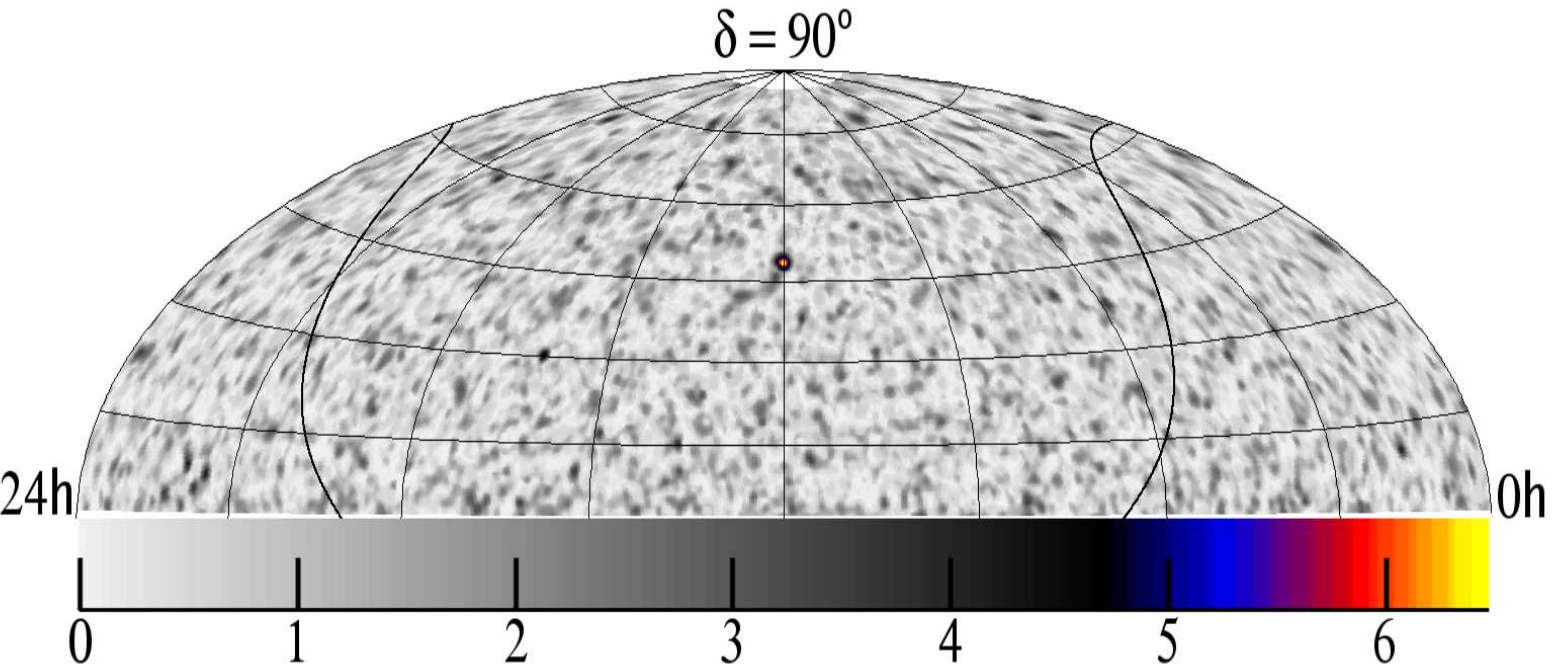}}
\caption{\label{Fig:Skymap} Significance sky map of 67 000 background events with
an added source of 15 events distributed according to E$^{-2}$ at declination $\delta$=48$^{\circ}$
and right ascension  $\alpha$=12h}
\end{center}\end{figure}

\begin{figure}\begin{center}
\begin{tabular}{cc}
\mbox{\includegraphics[width=2.57in]{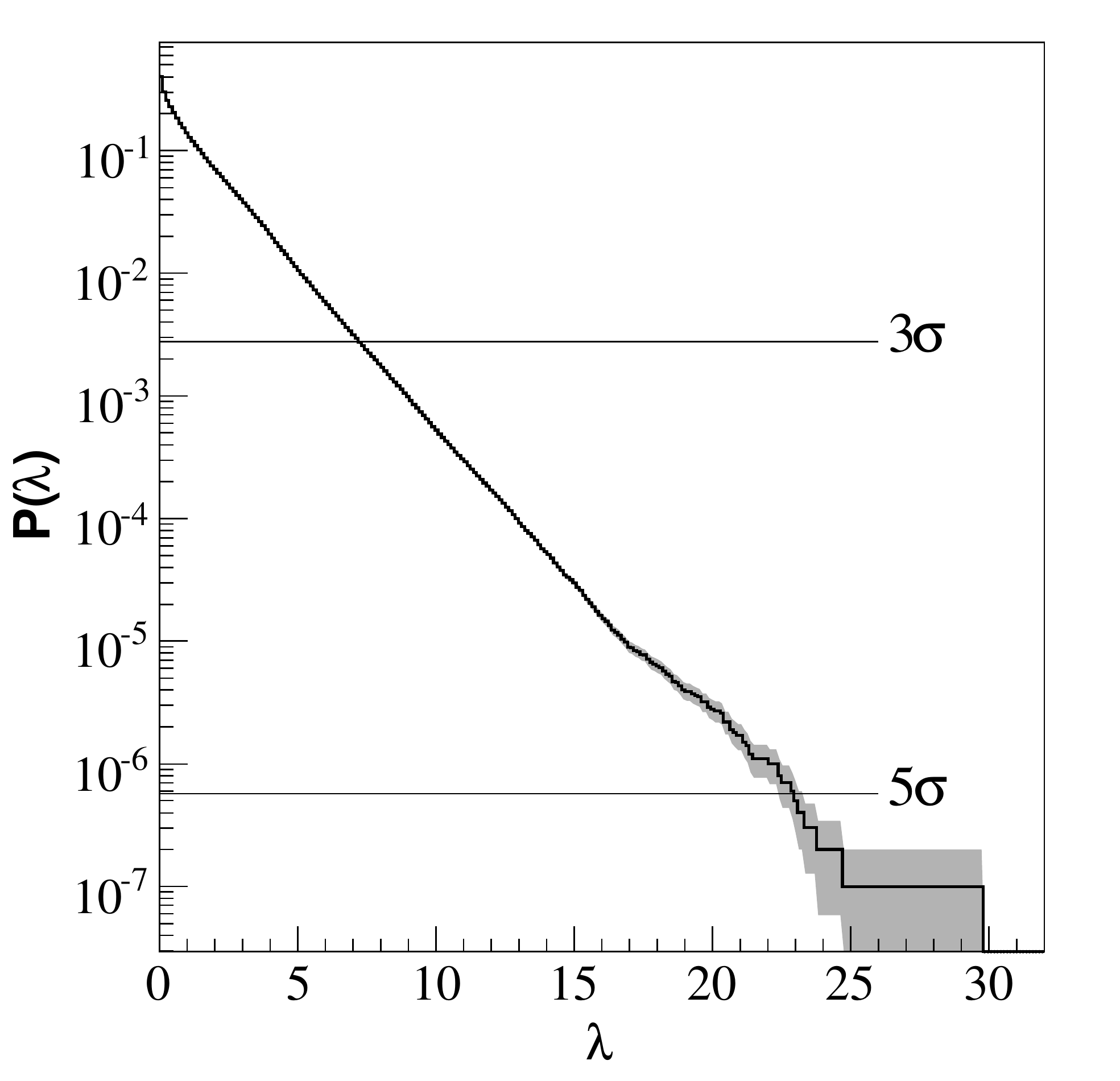}}
\mbox{\includegraphics[width=2.57in]{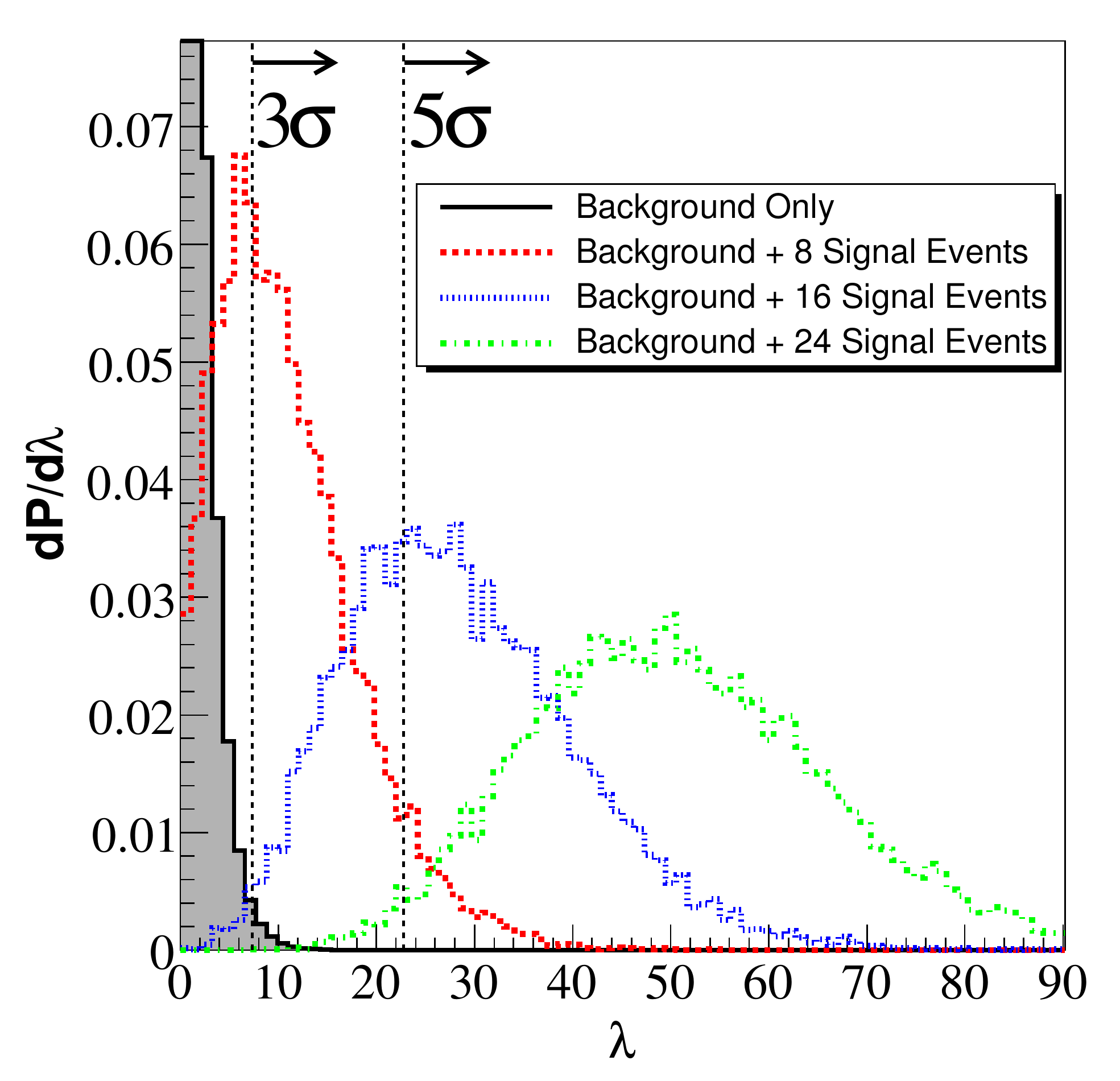}}
\end{tabular}
\caption{\label{Fig:Pdist1}  Left: Integral $\lambda$ probability distribution for background at declination $\delta$=48$^{\circ}$.
Shaded regions show statistical uncertainty.
The 2 horizontal lines indicate the values of the probabilities corresponding to $3\sigma$ and $5\sigma$ confidence levels.
Right: Distribution of $\lambda$ for background at declination $\delta$=48$^{\circ}$ and with 8, 16, and 24 added
signal events distributed according to an E$^{-2}$ energy spectrum.}
\end{center}\end{figure}

\begin{figure}\begin{center}
\mbox{\includegraphics[width=4in]{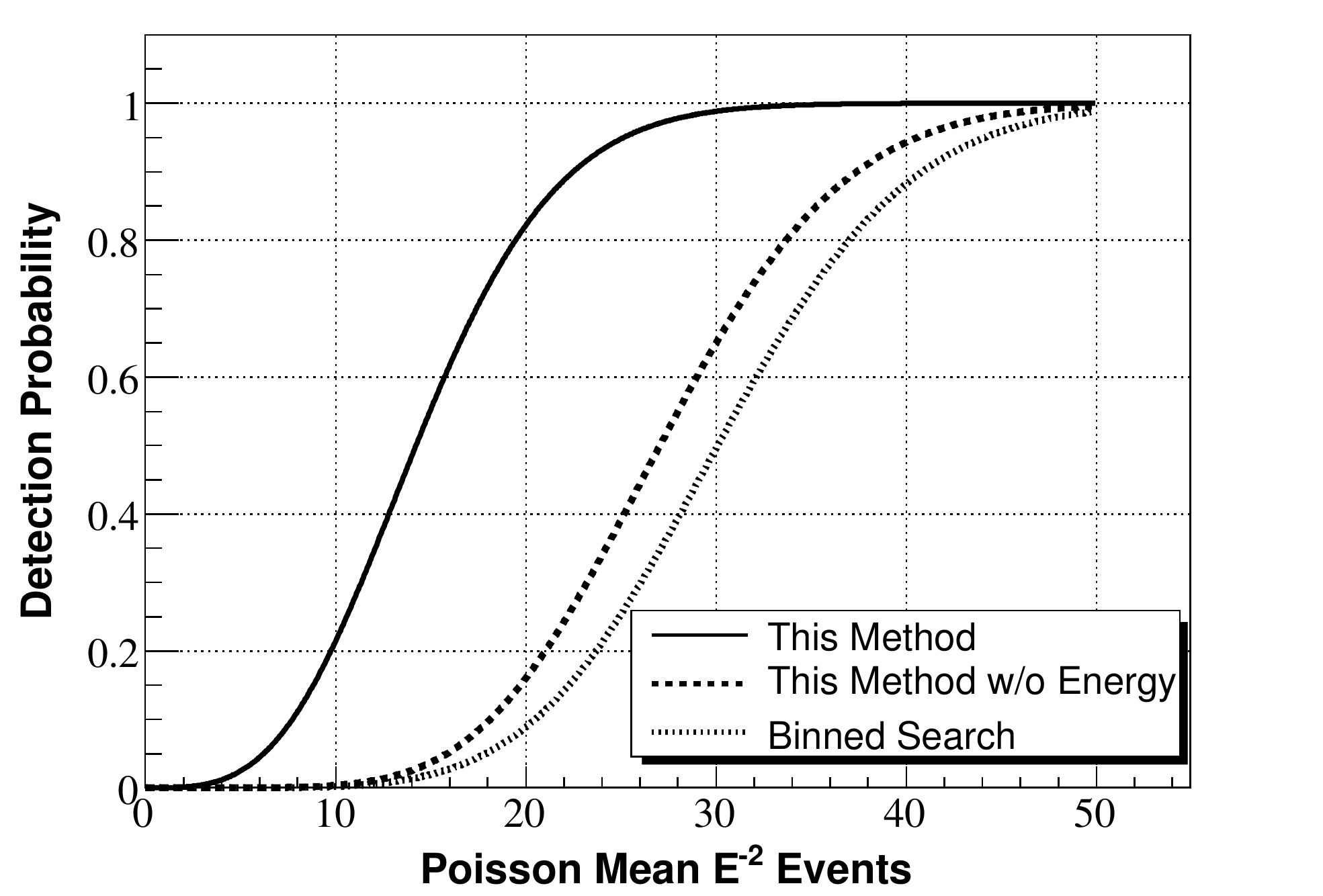}}
\caption{\label{Fig:Detprob} 5$\sigma$ confidence level detection probability vs. Poisson mean number of source
emitted events added to the background sample of 67 000 events, representing one year of data from a km$^3$ scale neutrino telescope.
The solid line indicates discovery potential for this method,
the dashed line for this method without using energy information, and the dotted line for the binned method.
The events follow an E$^{-2}$ energy spectrum. }
\end{center}\end{figure}

Fig.~\ref{Fig:Skymap} shows a significance sky map with an added source at declination $\delta$=$48^{\circ}$ and
right ascension $\alpha$=12h producing 15 events according to an E$^{-2}$ energy spectrum.
Fig.~\ref{Fig:Pdist1} illustrates the procedure used to determine significance and discovery potential.  The integral distribution
of $\lambda$ for background alone is produced at declination $\delta$=48$^{\circ}$ and the values of $\lambda$
corresponding to 3$\sigma$ ($2.7 \times 10^{-3}$)
and 5$\sigma$ ($5.7 \times 10^{-7}$) integral probability are calculated.  Fig.~\ref{Fig:Pdist1} also shows distributions
of lambda with 8, 16, and 24 signal events added to background.  Discovery potential at 5$\sigma$ is the fraction
of experiments with $\lambda$ exceeding the 5$\sigma$ threshold.  Discovery potential is then computed in this fashion
for each number of signal events and each spectral index at 5$\sigma$ confidence level.  The number of detected events
produced by a source of a given strength are Poisson distributed around a mean strength related directly
to the source flux.  The detection probability for a given mean strength is calculated by summing over the detection
probabilities for all 0 - 100 signal events and weighting by the appropriate Poisson probability.  The
detection probability at 5$\sigma$ confidence level as a function of mean source strength for an E$^{-2}$
signal flux is shown in Fig.~\ref{Fig:Detprob}.
As can be seen, the likelihood method requires approximately half the signal flux needed by the binned method to reach a similar
detection probability. To reach the same detection probability without using energy, the likelihood search still requires
$\sim$10\% less signal flux than the binned method.

\begin{figure}\begin{center}
\mbox{\includegraphics[width=4in]{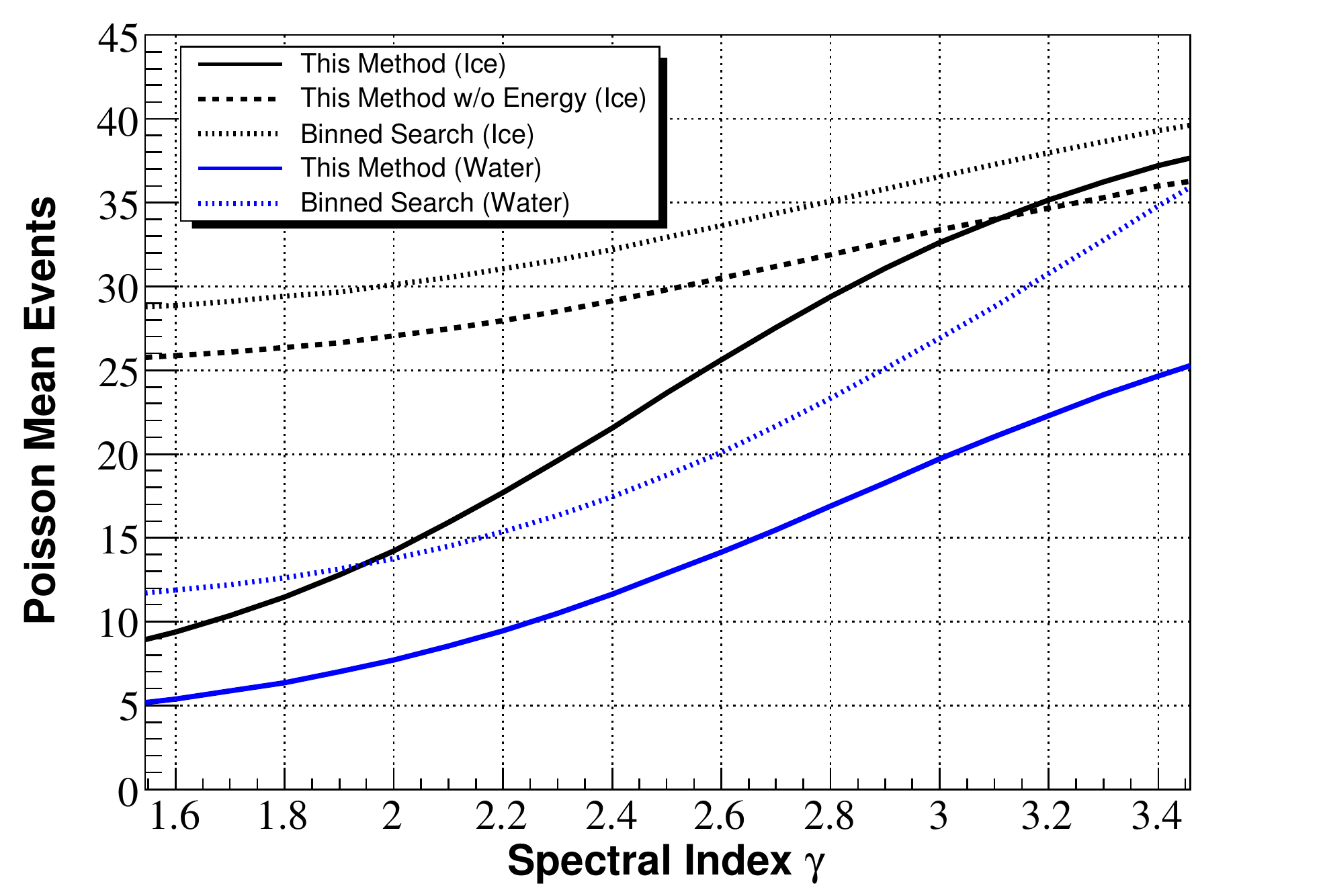}}
\caption{\label{Fig:Probvspec} Poisson mean number of signal events required for 50\% detection probability at
5$\sigma$ confidence level vs. source spectral index for the unbinned method described in the paper using
energy information in water (solid lower line) and in ice (solid upper line), not using energy information
in ice (dashed line) and for a binned search for a detector in water (dotted lower line) and in ice (dotted upper line).
For the detector in ice we assumed an angular resolution of 0.7$^{\circ}$, and we assume a resolution of 0.2$^{\circ}$ in water.
For both detectors, a background of 67 000 events is used.}
\end{center}\end{figure}

In Fig.~\ref{Fig:Probvspec} we show the 50\% detection probability at 5$\sigma$ confidence level as a function of source
spectral index for the three methods along with the values obtained from a similar analysis
with a detector capable of a muon angular resolution of $0.2^{\circ}$, similar to the resolution possibly achieved
by km$^3$ detectors in water \cite{ANTARES,distefano}.
With an angular resolution 0.2$^{\circ}$, the neutrino-muon vertex angle can no longer be neglected in the likelihood
method.  The resolution of 0.2$^{\circ}$ is convoluted with the neutrino-muon vertex angle as a function of 
reconstructed muon energy for an E$^{-2}$ neutrino spectrum in bins of 0.2 in log$_{10}$E.  The convoluted
resolution is used to compute $\mathcal{N}_i(\vec x_i|\vec x_s)$ in the likelihood method.  The bin size used by the binned method in water
is reoptimized and found to be 0.45$^{\circ}$.  Both detectors, ice and water, are simulated at the South Pole
to avoid differences in visibility as a function of declination.

For a source with an E$^{-2}$ energy spectrum, the binned method requires 30 signal events on top of background
for a 50\% chance of 5$\sigma$ detection, while the likelihood method requires 26 events without using
energy or only 14 events with energy.
For all methods, harder spectral indices require fewer events for 5$\sigma$ detection.  This trend is
caused by two factors:  Higher energy events have a smaller neutrino-muon vertex angle and thus better
resolution, and high energy astrophysical neutrinos produce muons at the detector with much higher energy
than typical atmospheric neutrinos.  The binned method is affected only by changes in the vertex angle,
while the likelihood method using event energy is affected by both factors and has a stronger response
to spectral index.  In ice, the method improves from only slightly better than binned methods for soft spectral
indices greater than $\sim$3 to more than a factor of two improvement for hard spectral indices less than $\sim$2.  A possible
way to improve the binned method is application of a cut on muon energy, keeping high energy events and
reducing background, but such a cut must be optimized assuming a specific source spectral index and is
generally undesirable.  Optimizing search bin size for a specific spectral index is similarly undesirable.
The likelihood method requires no a priori assumption of signal spectral index to utilize event energy
in discriminating signal from background.

Assuming a similar atmospheric neutrino background, the number of events required for 5$\sigma$ discovery in
sea water is a factor of $\sim$2 less than for ice if a resolution of 0.2$^{\circ}$ is achieved.  For a detector
in the Mediterranean, more than a year of livetime may be required to reach this background rate for declinations
with less than 100\% exposure.  With high statistics, significance improves inversely
with resolution, so one might expect a detector in ice with a resolution of 0.7$^{\circ}$ to require a factor of
3.5 more events than one in sea water with a resolution of 0.2$^{\circ}$ to achieve a similar significance.
However, in this analysis the factor
is smaller because event energy discriminates signal from background, and background rates are relatively low.

\begin{figure}\begin{center}
\begin{tabular}{cc}
\mbox{\includegraphics[width=2.57in]{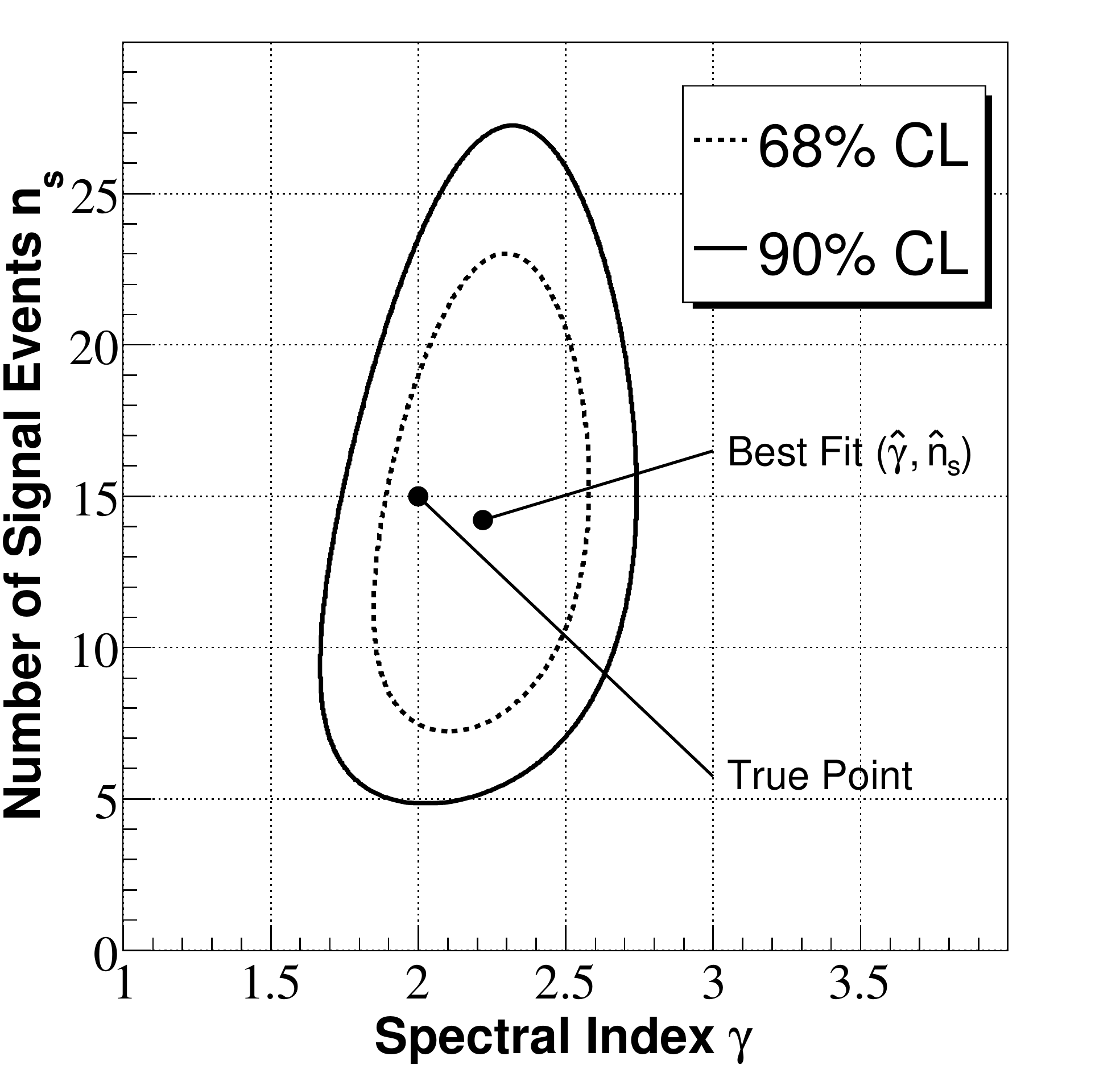}} &
\mbox{\includegraphics[width=2.57in]{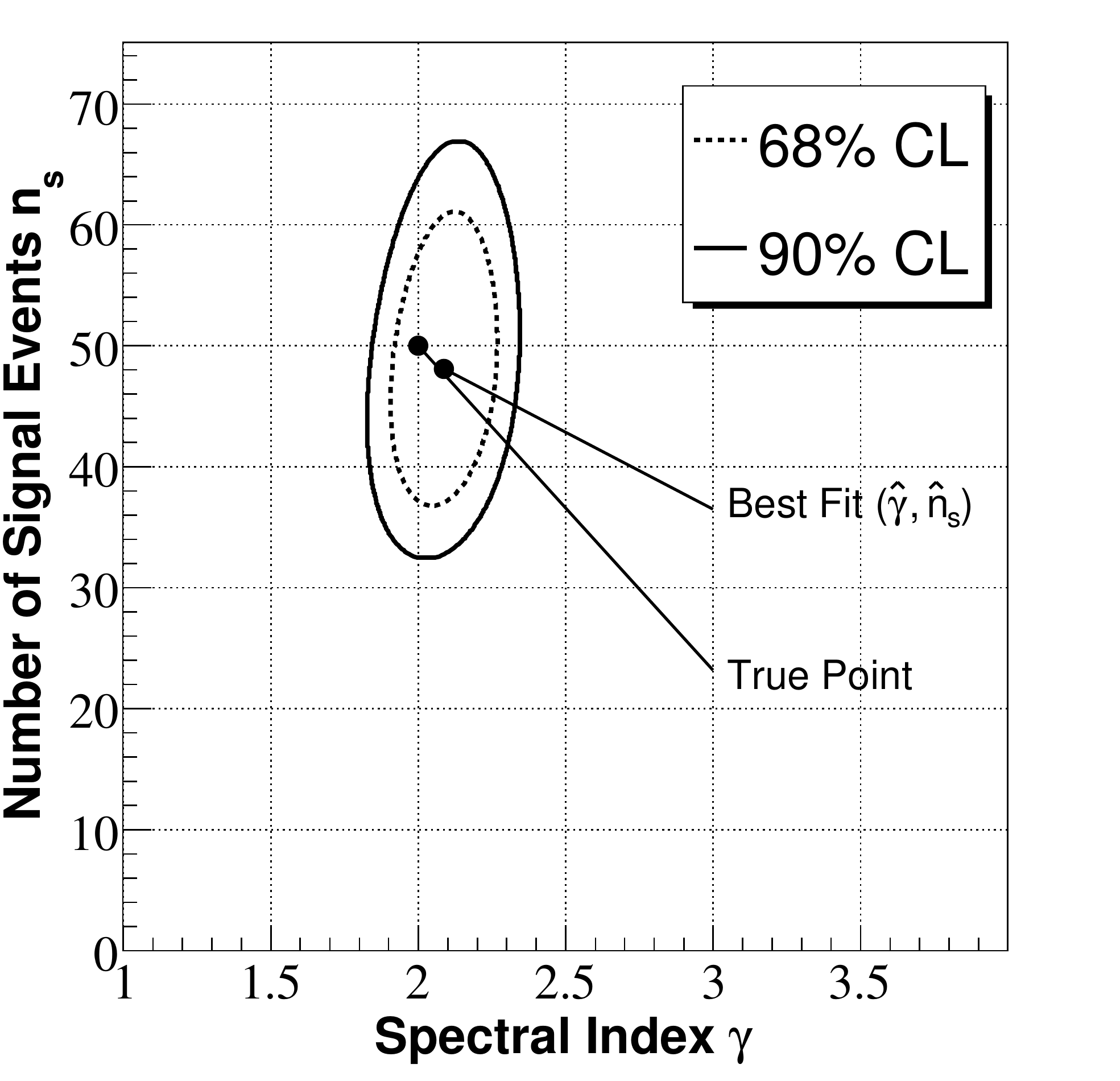}} \\
\end{tabular}
\caption{\label{Fig:Specest} Confidence boundaries are shown in  the spectral index vs. source strength parameter space
for 15 signal events (left) and 50 signal events (right) for 67\% and 90\% confidence levels.}
\end{center}\end{figure}

Finally, since source spectral index is a free parameter in the method and fitted to the most likely value,
the method provides an estimate of the spectral index.  Fig.~\ref{Fig:Specest} summarizes this capability for
source intensities of 15 events and 50 events following an E$^{-2}$ energy spectrum at declination $\delta$=48$^{\circ}$.
The distribution of
$-2 \cdot log[\mathcal{L}(n_s, \gamma)/\mathcal{L}(\hat{n}_s, \hat{\gamma})]$ is approximately given by a chi square
with two degrees of freedom.  Confidence contours are produced by comparing $-2 \cdot log$ likelihood ratio values
for points on the plane of source strength and spectral index to chi square values corresponding to 67\% and 90\% confidence level.
The coordinates of the best fit (the largest value of $\lambda$) and true point are indicated.  
Naturally, the confidence regions shrink for larger signal strengths, indicating better estimation of both parameters.
Good spectral index reconstruction is achieved with few events from a source despite limited energy resolution on an event by
event basis. This is possible because neutrino telescopes are capable of detecting neutrinos over many energy decades, offering a large
lever arm to discriminate energy spectra.

For a detector at the South Pole, background event density and angular resolution do not strongly depend on declination, 
so the results shown at $\delta$=48$^{\circ}$ generalize for declinations above $\delta$=0$^{\circ}$.
A detector in the Mediterranean will have stronger declination dependence of background event density due to
exposure differences, so the number of events required for 5$\sigma$ detection becomes smaller for declinations
with reduced exposure.

\section{Conclusions}
\label{Sec5}

We have shown that likelihood methods improve discovery potential using differences in both the angular distribution of events
and energy spectra between the background and signal hypotheses.  Furthermore, it has been shown that 10\% less flux is required
compared to methods using predefined angular search bins. The improvement is more significant when an energy related variable is
used to distinguish hard astrophysical neutrino spectra from the softer spectrum of atmospheric neutrinos. In this
case, for E$^{-2}$ energy spectra only about half of the flux is required to achieve 5$\sigma$ compared to binned methods.

\end{document}